# Securing the Network for a Smart Bracelet System

Iuliana Marin[1]
[1]Faculty of Engineering in Foreign Languages
University POLITEHNICA of Bucharest
Bucharest, Romania
marin.iulliana25@gmail.com

Nicolae Goga[1,2]
[2]Molecular Dynamics Group
University of Groningen
Groningen, Netherlands
n.goga@rug.nl

*Abstract*—Digital instruments play a vital role in our daily life. It is a routine to produce business papers, watch the news program, write articles and blogs, manage healthcare systems, to purchase online, to send messages and all this is processed by making observations and then manipulating, receiving and availing the diverse data. This electronic data provides the foundation of real time data. All this transmission of data needs to be secured. Security is essential for healthcare systems as the present one where the blood pressure recordings provided by the smart bracelet are sent to the user's mobile phone via Bluetooth. The bracelet monitors the pregnant women, but also other users who wish to have their blood pressure under control. The system's server analyses the recordings and announces the user, as well as the associated persons to the user in case of an emergency. The doctors, the medical staff, user and user's family and caregivers have access to the health recordings belonging to the monitored user. Security is a main feature of the electronic healthcare system based on the smart bracelet.

*Keywords—security, bracelet, blood pressure, healthcare, preeclampsia*

## I. Introduction

These days, the society is an information-intensive society. Researchers are engaged in creating, getting and utilizing various kinds of information during the course of our lives and, it is seen in most of the cases that this information takes the form of electronic data.

With the appearance of more electronic devices which are connected to the Internet, data privacy issues are starting to rise. In the era of fast communication, it is easier to generate and publicize data through the Internet, and hence data privacy problems get worse, especially in the real-time processing of large data. Data transmission real-time synchronization is required. It is needed to provide a strict protection for data privacy, and to require a compelling system architecture and computing power.

The most common examples of devices include computers and mobiles phones which deal with transforming the information into electrical data. For instance, approximately 500 million posts are being posted daily and 3.5 terabytes of the data are generated annually by the use of one of the renowned social networking site, Twitter [1]. IBM stated that today, most of the 90% of the data has been generated in the last two years. Therefore the need of information security is a must [2].

Field of study that deals with information protection mechanisms to ensure a level of confidence in this is called information security [3]. Information security comprises a defense in depth [4] that contains key elements such as the physical security solutions, the network, the hosts, the application and data.

The physical security solutions that protect data (locks, physical access control, monitoring). This is a very important layer as close to 70% of successful compromises happen because of poor implementation/lack of attention to this layer.

The network represents the architecture choices taken in order to secure the network of interconnected devices that host/process the data. The points of focus are the installation of network monitoring and securing devices, such as the instruction detection systems (IDS), intrusion prevention systems (IPS), firewalls.

Secure communication protocols for communication over a trusted or untrusted network are composed of the secure socket layer (SSL) and transport layer security (TLS). The host represents the security measures implemented for each individual host or machine, namely access control, user management, antivirus solution.

The application represents the security measures implemented within it that need to handle data for storing or processing. These measures are validation or sanitization of input, logging activity, authentication, error handling. Data represents the resource that information security is trying to protect. The data needs to be checked for validity and have copies or backups in order to be protected.

The Confidentiality, Integrity and Availability (CIA) triad supports information security [5]. Confidentiality deals with privacy of data and protecting it from being viewed by unauthorized third parties. Integrity deals with maintaining the validity of data based on accuracy, completeness and preventing unauthorized third parties from modifying it. Availability deals with problems that affect the capacity of requesting resources from a system, be them software problems such as attacks, bugs, errors or physical problems like hardware failures, power failures.

No functional device is 100% secure. The main goal of security is have the attacker cost higher than the value of the protected assets, while this value to be greater than the cost of the security assets [6].

Cyber-attacks are a daily occurrence and everyone is a potential target. Although the degrees of complexity between attacks may vary drastically, most attackers go for the Low Hanging Fruit, attacks that target recent vulnerabilities or easy to exploit vulnerabilities [7]. These attacks are launched against any and all machines that are connected to the Internet.

Therefore, the idea that a business is not worth attacking or will never be targeted is not trustful.

It is a common misconception that security devices solve all the problems, when, it truth, security devices are just part of the solution [8]. In actuality, people solve problems, in this case the core security team solves the problems, identify risks, determine what measures to be taken and install, deploy and maintain security devices.

The current paper aims secure the proposed network prototype. In the next section are presented the top used electronic healthcare system, along with how the patient's data is stored, the use of semantical annotations for such systems and the importance of personal details security is outlined. Section 3 outlines the description of the electronic healthcare system for monitoring the blood pressure and the prototype network architecture with its security. Section 4 presents the SSH brute force attack performed for the current electronic healthcare system. The last section presents the conclusions and the future work.

## II. Related Work

The traditional healthcare system gathers approximately 180 pages for each patient [9]. Nowadays, due to the electronic healthcare records systems doctors and medical practitioners keep the track of information related to the patient's health and offer access to the records through a centralized electronic system.

eClinicalWorks is the most used electronic healthcare system and it is very popular amongst neurology practices [10]. The security of this system consists in the determination of the permissions belonging to the users which access the program and the records of the patients. The administrator of the system has access to log files which offer information about the activity and that changes which were done [11].

McKesson has the oldest experience in the healthcare domain in America [12]. While most companies use as security measures firewalls, spam filters for the email, antivirus, McKesson considers that employees play a significant role in detecting doubtful emails and phone calls [13].

Cerner has the largest set of features in the industry of healthcare [14]. The security of the healthcare system offered by Cerner depends on the network design and deployment, fiber network, network security, as well as network monitor and management.

GE Healthcare offers hardware and software solutions to sustain healthcare facilities [15]. The provided security solutions incorporate infrastructure design and evaluation, lowering the risks, and monitoring the critical parts. The security products and services aim to diminish dangers, empower safe sharing of information to enhance patient care and fulfillment, along with ensuring data integrity.

Another solution is the use of HL7 standard for the management of the mother and child health [16]. The security for the electronic maternal and child heath registries is done through the usage of passwords for obtaining data, encryption of data which transits, storage of data independently from unique code identifiers [17].

The patient's data is stored according to the Clinical Data Interchange Standards Consortium Operational Data Model (CDISC ODM) which is based on the eXtensible Markup Language (XML) [18]. ODMedit is a web tool used to create data models based on semantic annotations [19]. This is done based on the metadata repository of medical data models belonging to experts [19].

Semantic annotation of consumer health questions was done for the email requested belonging to the U. S. National Library of Medicine customer service and from the questions posted on MedlinePlus [20]. After performing the annotation of the questions corpus, the results are used for machine learning methods in order to understand the content.

The automatic natural language processing (NLP) groups all research and development aimed to modeling and reproduce with the help of machines the human capacity to produce and understand linguistic statements for communication purposes. This processing has as key elements linguistics and computer science. It maintains close links with the cognitive science and has overlapping areas with Artificial Intelligence.

The part-of-speech (POS) tagging is done for each word. NN, NNS, NNP, NNPS are for nouns at singular or plural form. VB, VBD, VBG, VBN, VBP, VBZ are used to mark several forms of verbs. JJ stands for adjective, IN for preposition or subordinating conjunction.

Linked to the past research [21] where an automatic evaluation of answers to open questions, as well as providing of feedback for further reading or joining suitable expert communities, the current paper proposes a system where the user can write about his/her current health state and using natural language processing it is determined whether the post regards just an illness issue or a complaint.

Until now there is no such system which monitors the evolution and treatment of preeclampsia, as well as hypo and hypertension. The electronic healthcare system contains a knowledge base and an inference engine based on which the similarity between the existent cases is done and the treatment is offered.

The doctors can improve the knowledge base through the use of annotations and add new effective ways for treating preeclampsia which appears during the last trimester of the pregnancy and is characterized by the presence of hypertension. This illness is one of the three main causes of maternal death [22].

The semantical annotation was done using WordNet is an open source lexical database where nouns, adjectives and verbs are grouped into sets of synonyms. Each set is connected to a different concept.

Compared to the existent healthcare systems, the security of the healthcare system takes into consideration multiple solutions.

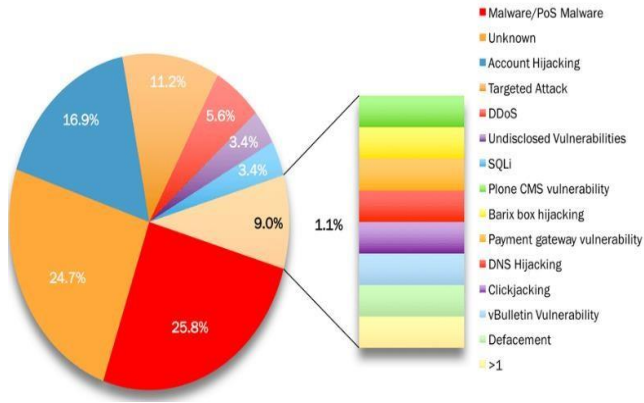

Fig. 1. Attack causes [23]

The security is done through the use of the proxy server, iptables, VPN connections, log traffic monitoring and classification, detection of malicious packets, encryption of data between the devices, as well as that coming from the Bluetooth module of the smart bracelet. The messages which are transmitted via the Bluetooth are encrypted using the Advanced Encryption Standard.

All the personal details about the patients, as well as the knowledge base need to be secured. According to the January 2017 cyber attacks statistics [23] (see Fig. 1), malware occupied the first position with a percentage of 25.8%, followed by account hijacking, target attack, distributed denial of service (DDoS), structured query language injection (SQLi), content management system (CMS).

The attacks targeted mostly the industry, government, education, healthcare, organization and military sectors [23]. Security measures need to be taken when dealing with research data coming not only from humans, but as well as animals, from which valuable information is extracted [24-26].

In 2017 the average cybercrime cost was of 17 million dollars for organizations in industries managing financial services, utilities and energy [27]. The percentage of increase in the cost of cyber security in a year is of 22.7% [27]. It takes on the average 50 days to solve a malicious insiders attack and 23 days to resolve a ransom ware attack.

## III. ELECTRONIC HEALTHCARE SYSTEM FOR BLOOD PRESSURE MONITORING

### A. System description

The health state of the pregnant women, as well as of the persons who wish to have their blood pressure monitored, have their recordings sent to their smart phone via Bluetooth. The mobile application has its own decision support and after processing the received value it displays the latest value of the blood pressure, as well as how it evolved. The recordings are also stored on the system's headquarters server [28, 29].

The doctors and the medical staff which are associated to the patient can view his/her electronic record and suggest a treatment through the use of annotations. The medical personnel collaborate in order to find out proper treatments which offer good outcomes. In this way, via annotations, the treatment can be improved. The doctors, the medical staff, the relatives and the caregivers associated to the user receive an SMS alert in case of a critical situation.

### B. Prototype Network Architecture and its Security

The network (see Fig. 2) is composed of the bracelet which transmits via Bluetooth the blood pressure recordings of the pregnant woman or of any other person who wants to have their blood pressure monitored in case of hypo and hypertension.

The messages transmitted by the Bluetooth module of the bracelet are encrypted using the Advanced Encryption Standard (AES). AES symmetric key algorithm depends on a symmetric key block cipher [30]. This encryption is more powerful than Triple DES. The calculations are done on the bytes of a matrix. Encryption comprises in byte substitution, row moving, mixing the columns and the addition of the round key. The decoding procedure comprises the addition of the round key, mixing of the columns, moving the rows, trailed by byte substitution.

The user can view his/her health state, as well as one of the persons who is associated to the user. The doctors and the medical personnel can view the details about the health state of their patients, as well as to assign and improve their treatment. All this data passes through the virtual private network (VPN) tunnel.

The computer acts as a reverse proxy server is a device on a private network that directs client requests to the appropriate backend, in this case, the NodeJS virtual private network (VPN) server or the CouchDB database server.

The Linux operating system of choice for implementing the servers, is Ubuntu 16.04. The installation is a minimalistic one, having initially installed only the Standard System Utilities, and SSH Server, the rest of the required packages being closely monitored and installed according to the role the server needs to play within the network.

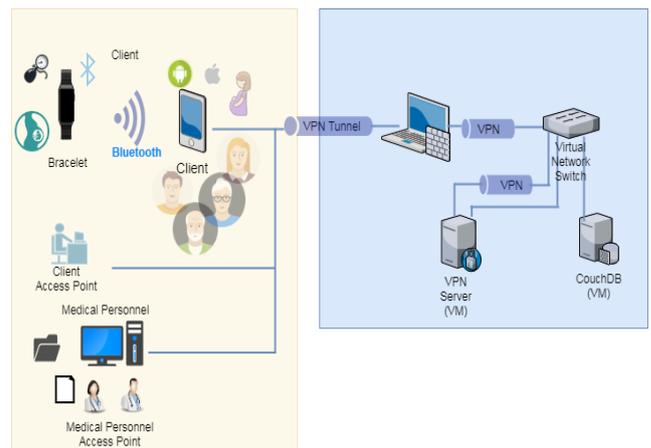

Fig. 2. Network prototype

The reasons for choosing this OS are the easy installation, the use Package Manager for installing and upgrading software, fast, easy on memory and highly modular, it has good system security, as well as for the security of the software. This OS is highly valued by security experts, developers and system administrators. In order to protect the server behind it, the proxy uses the iptables firewall, the ModSecurity web application firewall and the secure sockets layer (SSL) encryption.

Iptables is a set of rules organized in a table structure that are used to controlor restrict access over a network to and from a machine. Though iptables is used to set, view and manage the rules, the actual implementation of the rules is done by Netfilter.

Netfilter is a set of hooks at the kernel level that monitor the network stack, and apply the above-mentioned rules to the traffic (accept, deny, drop, log, masquerade, etc.).

The iptables are used in all devices to set FILTER rules, that only allow necessary services to be accessed by specific users, as well as network address translation (NAT) rules in specific devices, such as the firewall, that needs to for forward VPN traffic from the public interface to the private IP of the VPN server. Netfilter is also used by two other security applications within the prototype, namely port scan attack detector (PSAD) and Fail2Ban.

PSAD is involved in order to log traffic and block users that are attacking the server with DOS attacks, port scanning or launching multiple exploits. It is a lightweight daemon that detects suspicious traffic, such as port scans and sweeps, backdoor activity, botnet command and control communications, denial of service attacks.

The traffic is classified into the above categories using signatures from Snort intrusion detection system (IDS) software. PSAD leverages iptables in order to log suspicious traffic and block IPs that have been detected to be the source of the suspicious traffic, turning the software from an IDS to an intrusion prevention system (IPS).

Fail2Ban scans log files from multiple applications (Apache, SSH, etc.) in order to find signs of brute force attempts. If a brute force attack is identified, Fail2ban can block the attacker IP using iptables to block all incoming or outgoing traffic from and to that IP.

For the implementation Fail2Ban is configured to persistently and permanently block offenders that have tried to brute force the respective machine. The iptables firewall filters unwanted or malicious packets on the two to four layers of the open systems interconnection (OSI) stack.

ModSecurity web application firewall with custom rule set files (.crs) is used to detect malicious packets on the seventh layer of the OSI stack. The SSL encryption between client and proxy is involved in order to guarantee the privacy of the transmitted information.

The NodeJS web server frontend provides the user with an intuitive web interface to call the backend that controls and monitors the blood pressure recordings of the system.

The CouchDB database's main task is to identify dangerous situations by applying rules to the received blood pressure readings, and triggering appropriate events. To guarantee the privacy of the user's data, the reverse Proxy in front of the NodeJS server has SSL certificates to encrypt traffic, the user is not allowed to connect directly to the bracelet recordings.

The user must connect via a forwarded VPN connection of the healthcare system. This measure prevents denial of service (DOS) and DDOS attacks that target the smart bracelet, prevents unauthorized devices accessing the bracelet recordings, stop sniffers (between the healthcare system and the bracelet) from finding out the IP of remote user devices, as well as making it more difficult for attackers to target the bracelet.

OpenVPN was used for the prototype network. This is an open source Virtual Private Network software. It creates virtual network on the VPN server machine. Association is done by giving a virtual IP to the connected clients.

The network traffic will be routed through the VPN server, instead of requiring opening a shell sessions and running commands from the remote device VPNs have many applications, but in this case, it is used to create secure encrypted connections to the private network within the electronic healthcare system.

Once a client device is connected using valid certificates to the VPN server, by sending traffic through the tunnel interface it is seen as a device on the same network as the other devices in the network of the electronic healthcare system.

GPG, or GnuPG, stands for GNU Privacy Guard, and is used to encrypt and sign data that is supposed to be communicated with devices over an untrusted network. Data is encrypted using a passphrase and signed with PGP certificates.

The advantage is that the passphrase and encrypted data are sent separately (usually through different communication means), to prevent attackers from gaining access to the data unless they have both pieces of the puzzle, the passphrase and encrypted data being useless by themselves.

In the prototype, GPG is used to encrypt the user required OpenVPN files and certificates, in order for them to be securely sent to the client and, only decrypted once the ID of the client has been verified and confirmed.

Nginx is a open source, high performance, multifunctional server software. Its main features are the HTTP server, reverse proxy and mail proxy. In this case, the Nginx is configured as a SSL reverse proxy for the remote CouchDB database.

Multiple security measures are implemented in the reverse proxy to protect the server(s) that hides behind it. It needs to be a security dedicated server (functionality separation/network cohesion).

The load of the server it protects needs to be reduced. It has a layer of physical or virtual separation between the OS of the reverse proxy and sever behind it, if the proxy gets compromised.

## IV. SSH Brute force Attack

The SSH brute force is a brute force attack targeting the Secure Shell (SSH) service running on a machine. The goal of this attack is to reputedly try to authenticate using common or custom usernames and passwords until one or more valid combinations are found. TCP-Hydra, a popular multiprotocol brute force tool, was used with some supplied a list of common Unix/Linux usernames and passwords.

The first step the attacker will take is to recognize the target device. The tool of choice is Nmap. The command is nmap -A -T4 -p 1-1000 192.168.25.100, where -A enables the operating system and the service version detection, -T4 specifies the speed of the scan (where 1 is slowest, 5 is fastest and can even be considered a DOS attack), -p 1-1000 specifies ports to be scanned, in this case 1-1000 because they are the most common ports important services run on, and 192.168.25.100 which is the IP of target device.

In green appear the ports which are open and run SSH. Nmap also finds additional useful information like the version of SSH server, and information regarding the host-keys with which the server identifies itself to the client.

If the device was not running the SSH service on any port, then the SSH brute force attack would not be possible, but as this is not the case, the next step of the attack is reached by running the command hydra -L users.txt -P passwords.txt 192.168.25.100 ssh -t 4 -f -v, where -L users.txt gives the user.txt file content as possible usernames for brute force, -P passwords.txt gives the passwords.txt file content as possible passwords for brute force, 192.168.25.100 is the target's IP, ssh is the protocol for brute force.

By default, it translates to port 22. -t 4 represents the number of parallel child threads that launch the brute force attacks. -f stops the first successful user-password pair found (can be removed in order to enumerate all possible users, but takes longer) and -v is a verbose option. The measures to deal with SSH brute force attacks are to set firewall rules such that the SSH service can only be accessed from trusted IPs (stop attacks before they happen).

Another measure is to use Fail2Ban or other brute force prevention methods to permanently/temporarily block IPs launching the attack (stop ongoing attacks). The SSH server can be moved from port 22 to a nonstandard port (for example port 22222).

Uncommon usernames and strong passwords can be used, or where possible the SSH key authentication can be utilized. These measures are in order of importance and effectiveness, the set of the firewall rules being the most important.

## V. Conclusions

The architecture and the development of a system is a very complex undertaking, and the engineers involved in the development of the system right from the planning phase, analysis and design, should know how their envisaged project will work and look like.

There should be proper planning at the management level and at the technical level, and all requisite resources should be gathered before the project commences. However, even before spearheading the project, adequate security studies should be conducted so that stakeholder involved in the project are motivated by the fact that their solution will change the society.